\documentclass[aps,10pt,twocolumn,prb,showpacs,floatfix]{revtex4-1}

\usepackage{pifont}
\usepackage{upgreek}
\usepackage{amsfonts}
\usepackage{amssymb}
\usepackage{amsmath}
\usepackage{graphicx}
\usepackage{dcolumn}   
\usepackage{bm}        
\usepackage{flafter}

\begin{document}

\title{Superconductivity in SnO: a Nonmagnetic Analog to Fe-based
Superconductors?}
\author{M.~K. Forthaus$^{1}$, K. Sengupta$^{1,\dagger}$, O. Heyer$^{1}$,
N.~E. Christensen$^{2}$, A. Svane$^{2}$, K. Syassen$^{3}$,
D.~I. Khomskii$^{1}$, T. Lorenz$^{1}$, and M.~M. Abd-Elmeguid$^{1}$}
\affiliation{$^1$II. Physikalisches Institut,
Universit\"at zu K\"oln, Z\"{u}lpicher Str. 77, D-50937 K\"oln, Germany}
\affiliation{$^2$Department of Physics and Astronomy,
Aarhus University, Bygning 1520, Ny Munkegade 120, DK-8000 Aarhus, Denmark}
\affiliation{$^3$Max-Planck-Institut f\"{u}r
Festk\"{o}rperforschung, Heisenbergstra{\ss}e 1,
D-70569 Stuttgart, Germany}

\begin{abstract}
We found that under pressure SnO with  
$\alpha$-PbO structure, the same structure as
in many Fe-based superconductors,
e.g. $\beta$-FeSe,
undergoes a transition to a superconducting state
for $p\gtrsim6\,\text{GPa}$ with a maximum
$T_\text{c}$ of $1.4\,\text{K}$ at $p=9.3\,\text{GPa}$.
The pressure dependence
of $T_\text{c}$ reveals a dome-like shape
and superconductivity disappears for
$p\gtrsim16\,\text{GPa}$.
It is further shown from band structure calculations
that SnO under pressure exhibits a Fermi surface 
topology similar to that reported
for some Fe-based superconductors
and that the nesting between the hole
and electron pockets correlates with the change of $T_\text{c}$
as a function of pressure.
\end{abstract}



\bigskip

\maketitle

The discovery of superconductivity
in the iron-based layered compound
La(O$_\text{1-x}$F$_\text{x}$)FeAs \cite{kamihara2008} has
led to an explosion of interest not only in experimental
and theoretical studies but also in the search for new
compounds belonging to the same family of
materials. Several other groups of Fe-based superconductors
have been found, e.g. \textit{A}Fe$_\text{2}$As$_\text{2}$
(\textit{A} = Ba, K) \cite{rotter2008}
and more recently $\beta$-FeSe
\cite{hsu2008,mcqueen2009}. Their common features are
layers formed by FeAs tetrahedra, the presence of electron and hole
pockets in $k$-space at the Fermi energy, $E_\text{F}$, and the proximity to a magnetic
state. Although the pairing
mechanism is still a matter of debate \cite{mazin2009pc,chubukov2009pc}, it is
generally believed that superconductivity in FeAs 
compounds is unconventional
with electron pairing possibly mediated by spin fluctuations
\cite{mazin2008,kuroki2009} or a direct electron-hole pair hopping
\cite{chubukov2009pc}. Among the different types of Fe-based
superconductors, the nonmagnetic binary compound $\beta$-FeSe
(\textit{T}$_\text{C}=\text{8\,K}$) is of particular interest
owing to its simple structure. $\beta$-FeSe has a tetragonal
anti-$\alpha$-PbO type structure (\textit{P}4/\textit{nmm}) composed of
stacked FeSe layers along the \textit{c}-axis. It is, thus,
natural to ask whether compounds with the simple $\alpha$-PbO
structure but which do not contain a transition metal (Fe) would exhibit superconductivity.
SnO with the $\alpha$-PbO-type structure 
is such a candidate,
being a diamagnetic semiconductor at ambient
pressure ($p$) with an electronic structure
resembling that of Fe-based systems
\cite{lefebvre1998,watson1999}.
Moreover, according to recent infrared
and x-ray diffraction measurements 
SnO becomes metallic at room temperature
above $p=5\,\text{GPa}$ without a change of the lattice
structure up to $p \sim17\,\text{GPa}$ \cite{wang2004}.
This finding of
a pressure-induced metallic
state of SnO above $\sim5\,\text{GPa}$
is in good agreement
with theoretical band structure
studies using
LDA calculations
\cite{christensen2005}.

In this work, we show
that SnO becomes superconducting
under pressure
$p\gtrsim6\,\text{GPa}$
with a maximum $T_\text{c}$
of 1.4\,K at
$p\sim9\,\text{GPa}$
and that
$T_\text{c}(p)$
reveals a dome-like
shape similar
to other Fe-based superconductors.
Our band structure
calculations indicate
that SnO exhibits
a Fermi surface
topology similar to that
reported for
Fe-based superconductors
and that the nesting
between the hole
and electron pockets
correlates with
$T_\text{c}(p)$.

\begin{figure}[htbp]
\centering
\includegraphics[angle=0,width=\linewidth]{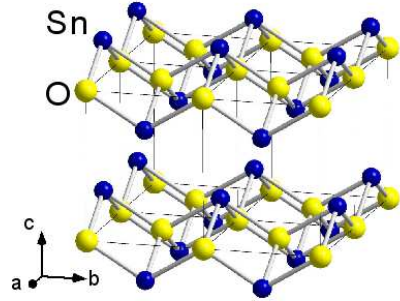}
\caption{(color online.)
Crystal structure of SnO
($\alpha$-PbO-type structure 
(\textit{P}4/\textit{nmm})).}
\label{figure0}
\end{figure}

For the four-probe
electrical resistivity
measurement
polycrystalline single-phase 
samples with a stated
purity of $99.9\,\%$
purchased from Alfa
Aesar were gently ground and
inserted into the
sample chamber of a diamond
anvil cell (DAC)
with a cell body made of
a special Ti-alloy
to minimize temperature-induced
variations of the pressure. 
Hardened Inconel\,750 was used
as gasket material,
electrical insulation 
of the gasket from 
the four gold leads 
and the sample was
provided by
a mixture of epoxy
and fine Al$_2$O$_3$ powder.
Pressure was determined
at room temperature using
the ruby fluorescence method
\cite{syassen2008}.
The temperature
dependence of the
electrical resistivity
$\rho(T,p)$ was measured
in a $^4$He bath cryostat
which covers the temperature
range $1.6\,\text{K}\leq T\leq300\,\text{K}$.
Measurements at low
temperatures ($0.4\,\text{K}\lesssim T\leq15\,\text{K}$)
and in external magnetic
fields
where performed in a
HelioxVL $^3$He refrigerator
(Oxford Instruments).


\begin{figure}[htbp]

\vspace{-0.3cm}

\includegraphics[angle=0,width=0.85\linewidth]{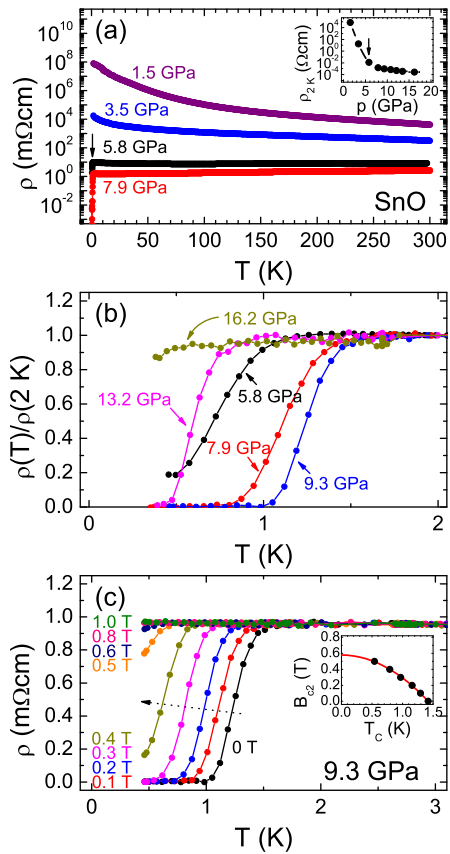}

\vspace{-0.6cm}

\caption{(color online). 
(a) Electrical resistivity $\rho(T,p)$ (logarithmic scale)
of SnO for $0.4\,\text{K}\lesssim T\leq300\,\text{K}$
and at different pressures up to 7.9\,GPa. The arrow indicates
the onset of the superconducting transition for $p\geq5.8\,\text{GPa}$.
The inset shows the pressure dependence of the
resistivity at 2\,K ($\rho_\text{2\,K}(p)$). 
The arrow at 5.8\,GPa marks the first observation of $T_\text{c}$.
(b) Low temperature region ($T\leq\text{2\,K}$)
of $\rho(T)/\rho(\text{2\,K})$
for different pressures from 5.8\,GPa to
16.2\,GPa. 
(c) Effect of an external magnetic field
$\text{0\,T}\leq B_\text{ex}\leq\text{1\,T}$
on $\rho(T)$ for 9.3\,GPa and
$T\leq\text{3\,K}$. The inset
shows $B_\text{c2}(T_\text{c})$, the red line corresponds to
a fit with $B_\text{c2}(\text{0})=\text{0.58\,T}$
(see text).}
\label{figure1}
\end{figure}

Fig. \ref{figure1}(a) displays the temperature dependence
of the electrical resistivity $\rho(T)$ in the temperature
range $0.4\,\text{K}\leq T\leq300\,\text{K}$ at 
selected pressures up to $7.9\,\text{GPa}$ 
using a diamond anvil cell. We find that
with increasing pressure the resistivity gradually changes
from the semiconducting to a metallic-like behavior,
followed by a sharp drop of $\rho(T)$ ($\sim80\,\%$ at
$0.35\,\text{K}$ at $5.8\,\text{GPa}$) at a critical temperature
$T_\text{onset}=1.3\,\text{K}$, indicative of the onset of a
superconducting transition.
The temperature of the superconducting transition increases
with further increase of the pressure to $T_\text{onset}=1.7\,\text{K}$
at $7.9\,\text{GPa}$ where the system exhibits zero resistivity at
$0.8\,\text{K}$. The proximity of the observed
superconducting transition at $5.8\,\text{GPa}$ to a
semiconductor to metal in SnO transition is demonstrated in
the inset of Fig. \ref{figure1}(a). Here, one observes for
$p\leq5.8\,\text{GPa}$ a dramatic decrease of the
value of $\rho(2\,\text{K})$, i.e. the resistivity in the normal state
just above the superconducting transition,
by more than 6 orders of magnitudes. This decrease becomes distinctly weaker
for $5.8\,\text{GPa}\leq p\leq16.2\,\text{GPa}$.
We note that the observation
of a metallic-like behavior at
$p = 5.8\,\text{GPa}$ is in a good agreement with
that reported from high pressure infrared spectroscopy
\cite{wang2004}.
The superconducting transition temperature $T_\text{c}$
for SnO for
$5.8\,\text{GPa}\leq p \leq16.2\,\text{GPa}$
can be seen in Fig. \ref{figure1}(b). Here, it is shown that the
superconducting transitions are relatively sharp up to the highest
pressure of $p=13.2\,\text{GPa}$, although we cannot
exclude the presence of some pressure gradients
in the sample chamber. At $16.2\,\text{GPa}$, we find an
onset of the superconducting transition at $\sim0.6\,\text{K}$, below which
$\rho(T)$ drops by $10\,\%$ down to $\sim0.4\,\text{K}$.
The values of $T_\text{c}$ are determined from the temperature
at which $\rho(T)$ is reduced by $10\,\%$ below $T_\text{onset}$,
i.e. $\rho(T_\text{c})/\rho(T_\text{onset})=0.9$.
The effect of an external magnetic field $B_\text{ex}$ on the
temperature dependence of the electrical resistivity at
$p = 9.3\,\text{GPa}$ which corresponds to the maximum value
of $T_\text{c} = 1.4\,\text{K}$ is shown in Fig. \ref{figure1}(c).
As expected for bulk
superconductivity, $T_\text{c}$ initially shifts 
linearly to lower temperatures
and remains sharp
with increasing $B_\text{ex}$ and can be described by the
Werthamer-Helfand-Hohenberg (WHH) theory for
{\textquotedblleft}dirty superconductors{\textquotedblright}
\cite{werthamer1966}. This excludes the existence of weak
links and/or minority phase superconductivity in the sample.
We obtain for the superconducting state of SnO at $p = 9.3\,\text{GPa}$ a value of the
upper critical field $B_\text{c2}(T = 0)$ of about $0.58\,\text{T}$ (see inset
of Fig. \ref{figure1}(c)) which corresponds to a coherence length
$\zeta\approx240\,$\AA.

\begin{figure}[htbp]

\vspace{-0.3cm}

\includegraphics[angle=0,width=0.85\linewidth]{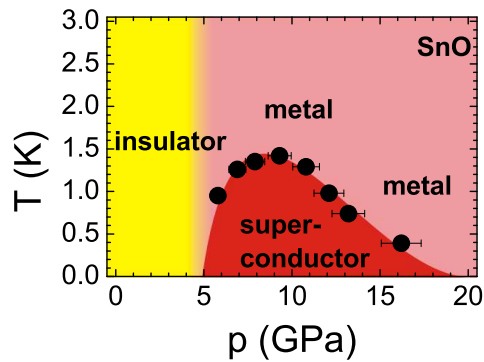}

\vspace{-0.6cm}

\caption{(color online). 
Electronic ($p,T$) phase diagram of SnO.
The semiconductor to metal transition
is around $\sim\text{5\,GPa}$. $T_\text{c}(p)$
shows a dome-like shape with 
a maximum ($T_\text{c}\sim\text{1.4\,K}$)
around 9\,GPa, and decreases gradually for
$\text{9\,GPa}\lesssim p\leq\text{16.2\,GPa}$
with $T_\text{c}(16.2\,\text{GPa})\sim0.4\,\text{K}$}
\label{figure2}
\end{figure}

The results of our high pressure investigation
of SnO are summarized in a $(p,T)$ phase diagram in Fig. \ref{figure2}. The pressure
dependence of $T_\text{c}$ displays a dome-like shape with
a maximum value of $T_\text{c}$ of about
$1.4\,\text{K}$ around $9\,\text{GPa}$.
$T_\text{c}$ increases with
increasing pressure with an 
initial rate of
$\partial T_\text{c}/\partial p\approx0.28\,\text{K}\,\text{GPa}^{-1}$
between 5.8\,GPa and 6.9\,GPa,
passes through a maximum around $9\,\text{GPa}$, and then
decreases towards zero for $p>16.2\,\text{GPa}$.
This behavior of $T_\text{c}(p)$
of SnO resembles that recently
reported from high pressure studies on $\beta$-FeSe
superconducting
samples \cite{medvedev2009,braithwaite2009}.
However, in $\beta$-FeSe the value of $T_\text{c}$
at ambient pressure is already much higher
than the maximum $T_\text{c}$ in SnO under high pressure. Also the initial
increase of $T_\text{c}$ with pressure is stronger for FeSe 
than we find for SnO in the superconducting
phase. On the other hand, in both cases the
initial increase of $T_\text{c}$ with pressure is
associated with a corresponding decrease of the
interlayer distance (Se--Se/Sn--Sn) along the \textit{c}-axis
\cite{wang2004,medvedev2009}.

\begin{figure}[htbp]

\vspace{-0.3cm}

\includegraphics[angle=0,width=0.85\linewidth]{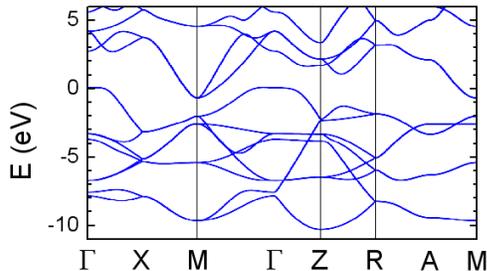}

\vspace{-0.6cm}

\caption{(color online). 
The band structure of SnO at 7\,GPa. The calculations
base on the structural parameters determined by x-ray diffraction
experiments \cite{wang2004}. The Fermi level is at $E=0$.}
\label{figure3}
\end{figure}

It is interesting to compare the
properties of SnO under high pressure
with those of Fe-based superconductors.
We recall the  main ingredients for
the occurrence of superconductivity in the latter systems:
(i) the existence of several pockets in the Fermi surface with some
specific topology; (ii) some degree
of nesting is required between the hole and
electron pockets; (iii) spin fluctuations associated with
such nesting seem to be important.
However, the relative importance of the above mentioned ingredients
for superconductivity is still a matter of debate. Here, we mention the
two most discussed but quite different scenarios. According to
Mazin \textit{et al.} \cite{mazin2008}
and Kuroki \textit{et al.} \cite{kuroki2009},
all three factors (i)--(iii) are essential. In this approach
superconductivity is induced by the nesting-related antiferromagnetic
spin fluctuations near the wave vectors connecting the electron
and hole pockets. This leads to an extended \textit{s}-wave pairing
with a sign reversal of the order parameter (\textit{s}$^\pm$) between
different (nested) Fermi surface sheets. In contrast to this,
according to Chubukov \cite{chubukov2009pc},
and Kuchinskii and Sadovskii \cite{kuchinskii2009}
predominantly (i) and (ii) are relevant. These authors
proposed, based on a weak coupling approach that
spin fluctuations are not necessarily required for an \textit{s}$^\pm$
state and the pairing mechanism might not be magnetically mediated.
It can predominantly originate e.g. from a direct pair hopping between
hole and electron Fermi surfaces. We note that
here the pockets are important but only nearly nested states are necessary.

In view of the two theoretical approaches mentioned above, it appears
that the existence of pockets and some degree
of nesting are relevant for the occurrence of
superconductivity in Fe-based superconductors. As SnO does not
contain magnetic atoms, we explore ingredients (i) and (ii) in the case of SnO.
We thus have investigated the Fermi surfaces and nesting properties of
metallic SnO and their change with pressure. Fig. \ref{figure3} illustrates the calculated
bands of metallic SnO at $7\,\text{GPa}$ along symmetry lines in the Brillouin zone (BZ).
These calculations are scalar-relativistic and 
use the Linear-Muffin-Tin-Orbital (LMTO) method \cite{andersen1975} 
in a full-potential implementation \cite{methfessel2000}. 
As shown in Fig. \ref{figure3},
the band structure close to $E_\text{F}$
is similar to that in FeAs systems \cite{mazin2008,subedi2008}:
the two bands dipping below $E_\text{F}$ near the 
M-point of the Brillouin zone enclose two
electron pockets centered at M whereas the rather flat band
slightly above $E_\text{F}$ around $\Gamma$ encloses a hole surface. 
The states near $E_\text{F}$ at M move further down relative
to $E_\text{F}$ as the pressure is increased, whereas the
{\textquotedblleft}hole band{\textquotedblright}
shifts towards higher energies. The indirect band overlap increases with pressure. The
{\textquotedblleft}hole band{\textquotedblright} states as
defined here have O-$p_\text{z}$ and Sn-$s$ character, whereas the
{\textquotedblleft}electron states{\textquotedblright}
close to $E_\text{F}$, around M are of Sn-$p_\text{x}p_\text{y}$
character (coordinates $x$, $y$, and $z$ are along the tetragonal axes,
$c$-axis in the $z$-direction).

\begin{figure}[htbp]

\vspace{-0.3cm}

\includegraphics[angle=0,width=0.85\linewidth]{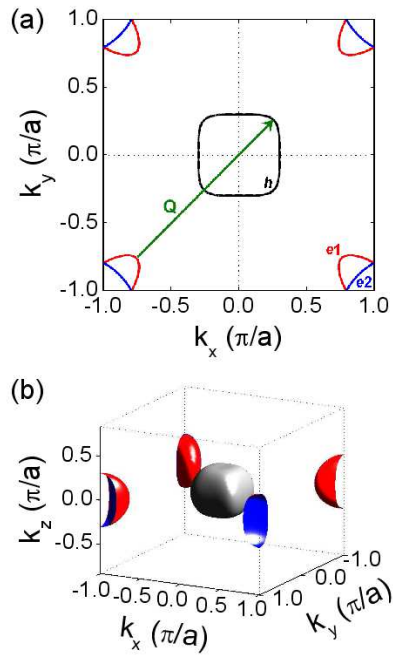}

\vspace{-0.6cm}

\caption{(color online). 
(a) Fermi surface contours in the ($k_\text{x},k_\text{y}$) 
plane as calculated for SnO at 7\,GPa. The blue and red curves at
the corners (M points) are contours of the electron surfaces,
e1 (outer surface, red) and e2 (inner surface, blue). The black
curve enclosing the BZ centre ($\Gamma$) is the contour of the hole,
h, surface. The vector $\textbf{Q}$ is a representative of nesting vectors
in the (1,1,0) directions between the e1 and h surfaces.
(b) Fermi surface for SnO
calculated at 7 GPa.}
\label{figure4}
\end{figure}

Figure \ref{figure4}(a) displays the Fermi surface contours in the
($k_\text{x}$,$k_\text{y}$) plane of the
BZ as calculated for SnO at $p=7\,\text{GPa}$.
We see that the band structure close to
$E_\text{F}$
and the shape of the Fermi surface
of SnO are
quite similar to those reported for $\beta$-FeSe \cite{subedi2008}
and other Fe-based superconductors.
The SnO band structure is, however,
nearly three-dimensional:
the Fermi surface does not exhibit pronounced cylindrical
sheets as in FeSe, but has a hole surface 
of box-like shape (see Fig. \ref{figure4}(b)). 
However, there is still an efficient nesting
between the outer electron surface (e1) and the hole surface (h),
see Fig. \ref{figure4}(a). The Fermi surface topology does not
change with pressure in the pressure range considered
here (5 to 19\,GPa), but the degree of nesting between
the electron and hole FS changes. 

\begin{figure}[htbp]

\vspace{-0.3cm}

\includegraphics[angle=0,width=0.85\linewidth]{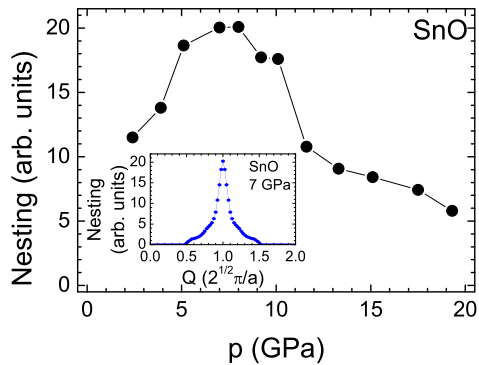}

\vspace{-0.6cm}

\caption{(color online.) The maximum value of the nesting functions vs. pressure.
The values of pressure, volumes, and structural parameters
are taken from the experimental values from Ref. 11.
The inset shows an example of the calculation 
of such a nesting function $N(\textbf{Q})$
for $p=7\,\text{GPa}$ (see text).}
\label{figure5}
\end{figure}

The variation of
nesting with pressure illustrated in Fig. \ref{figure5} has been
calculated from the maximum values of the nesting
functions for the structural parameters at different
pressure points \cite{wang2004}. An example of the calculation 
of such a nesting function $N(\textbf{Q})$,
which is given by
$N(\textbf{Q})=\int_\text{BZ}\text{d}^3k\,\delta(E_\textbf{k}-E_{\textbf{k}+\textbf{Q}})\delta(E_\textbf{k}-E_\text{F})$ 
is shown in 
the inset of Fig. \ref{figure5}. 
As evident from
Fig. \ref{figure5}, the degree of nesting first increases with increasing
pressure, reaches a maximum between 8 and 10\,GPa
and decreases with further increasing pressure. This
behavior correlates with the observed pressure
variation of $T_\text{c}$ (see $(p,T)$-phase diagram; Fig. \ref{figure2}).
In contrast to this, the
calculated
density of states at $E_\text{F}$, 
$D(E_\text{F})$,
as a function of pressure reveals
no remarkable change up to
about 20\,GPa: while 
$T_\text{c}$ strongly decreases
by more than a factor of 3 
between 9\,GPa and 16\,GPa,
$D(E_\text{F})$ shows only
a slight decrease of about
10\,\% in the same pressure range.
These theoretical
calculations demonstrate that even in the
absence of spin fluctuations
the Fermi surface
topology and nesting properties are relevant
to the observed
pressure-induced superconductivity
in SnO .
In such a case the
observed superconductivity
in metallic SnO with
much lower $T_\text{c}$ than
that found in Fe-based superconductors
may be explained in the
weak coupling scenario
by direct pair hopping
between electron and hole pockets
\cite{chubukov2009pc}.
In contrast, spin fluctuations usually
lead to strong electronic correlations,
i.e. strong coupling and,
thereby, higher values of $T_\text{c}$.

In summary, we discovered that SnO, with 
the $\alpha$-PbO-type structure, undergoes 
an insulator-metal transition under pressure 
at $p_\text{c}\sim6\,\text{GPa}$ and becomes 
superconducting with $T_\text{c}\sim0.4\,\text{K}$. With 
increasing pressure, $T_\text{c}$ initially 
increases up to a maximum of 1.4\,K at  
9.3\,GPa, but then decreases and 
superconductivity disappears at $p\sim16\,\text{GPa}$. 
The electronic structure of SnO is that 
of a semi-metal with the hole pocket at 
the $\Gamma$-point and electronic pockets at 
M-points of the Brillouin zone. Thus, both 
the crystal and electronic structures of 
SnO are similar to those of Fe-based 
superconductors like FeSe or FeAs-systems, 
but in contrast to these SnO is nonmagnetic, 
and the values of $T_\text{c}$ and $B_\text{c2}$ 
($B_\text{c2}\sim0.6\,\text{T}$  at 9.3\,GPa) 
are much smaller. We believe that, 
besides its own interest, the discovery of 
superconductivity in SnO, the nonmagnetic 
counterpart of Fe-based superconductors, may 
be useful for a better understanding of the 
relative importance of different factors relevant 
for superconductivity in this class of materials.

This work was supported
by the Deutsche Forschungsgemeinschaft 
(DFG) through SFB 608.
K.S. and M.M.A. would like to thank
the Alexander von
Humboldt foundation for the 
financial support.

\end{document}